%% file: main.tex
\documentclass[acmsmall,screen,nonacm]{acmart}

\usepackage{booktabs} % For formal tables
\usepackage{listings}
\usepackage{xcolor}
\usepackage{balance}
\usepackage{mathpartir}
\usepackage{amsmath, amssymb} % Standard math packages
\usepackage{enumitem} % For finer list control if needed (optional)

\begin{document}

\title{Extending Data Spatial Semantics for Scale Agnostic Programming}

\author{Jason Mars}
\affiliation{
  \institution{University of Michigan }
  \streetaddress{2260 Hayward St.}
  \city{Ann Arbor}
  \state{Michigan}
  \country{USA}
  \postcode{48104}
}
\email{profmars@umich.edu}

\input{0-abstract}

\maketitle

\input{1-intro}

\input{2-casefor}

\input{3-semantics}

\input{4-implications}

\input{5-implementation}

\bibliographystyle{ACM-Reference-Format}
\bibliography{references}

\appendix

\onecolumn

\end{document}

%% file: 0-abstract.tex
\begin{abstract}
We introduce extensions to Data Spatial Programming (DSP) that enable \emph{scale-agnostic programming} for application development. Building on DSP's paradigm shift from ``data-to-compute'' to ``compute-to-data,'' we formalize additional intrinsic language constructs that abstract persistent state, multi-user contexts, multiple entry points, and cross-machine distribution for applications. By introducing a globally accessible root node and treating walkers as potential entry points, we demonstrate how programs can be written once and executed across scales—from single-user to multi-user, from local to distributed—without modification. These extensions allow developers to focus on domain logic while delegating runtime concerns of persistence, multi-user support, distribution, and API interfacing to the execution environment. Our approach makes scale-agnostic programming a natural extension of the topological semantics of DSP, allowing applications to seamlessly transition from single-user to multi-user scenarios, from ephemeral to persistent execution contexts, and from local to distributed execution environments.
\end{abstract}

%% file: 1-intro.tex
\section{Introduction}
\label{sec:introduction}

Programming language design has long focused on providing abstractions that shield developers from underlying implementation complexities. At their core, successful abstractions identify patterns that are both ubiquitous and amenable to standardization.

Certain concerns arise so frequently in modern software development that they warrant representation as intrinsic language constructs rather than library-based extensions. While concepts like data structures, control flow, and computation have earned their place as fundamental language elements, other equally ubiquitous concerns remain relegated to libraries, frameworks, or operating system interfaces. This creates an artificial separation between domain logic and operational concerns that complicates development, particularly as applications scale.

Data Spatial Programming (DSP)~\cite{mars2025dataspatialprogramming} introduced a paradigm shift from moving data to computation toward moving computation to data through topologically-aware constructs. DSP formalizes these topological relationships within the programming model through archetypes like nodes, edges, and walkers. This enables more natural representations of computational problems as graph-like structures, dynamic traversal patterns, and context-dependent behaviors.

However, DSP's core formulation, along with most programming paradigms, does not directly address several ubiquitous concerns in modern application development:

\begin{enumerate}
    \item \textbf{Persistence across executions:} Applications increasingly require state preservation between runs, yet persistence mechanisms remain external to most language semantics.
    
    \item \textbf{Multi-user contexts:} Modern applications typically serve multiple users with isolated state, but multi-user semantics are rarely expressed directly in programming models.
    
    \item \textbf{Execution models beyond single-entry, single-exit:} While service-oriented and event-driven architectures dominate modern application design, language constructs often retain a bias toward batch-processing execution models.
    
    \item \textbf{Distribution across machines:} As applications scale, computation often needs to span multiple machines, yet distribution remains largely external to language semantics, requiring significant adaptation of single-machine code.
\end{enumerate}

We argue that these concerns should be elevated to intrinsic language constructs in programming language design, particularly for paradigms like DSP that already formalize topological relationships. By extending DSP semantics with constructs that directly represent persistence, multi-user contexts, multiple entry points, and cross-machine distribution, we can create a what we term a \textbf{scale-agnostic programming} model where applications written at the individual scale naturally extend to multi-user and distributed contexts.

In this paper, we present extensions to DSP semantics that support scale-agnostic programming through three key constructs: a persistent root node accessible throughout the program, the treatment of walkers as potential entry points into long-running applications, and topology-aware distribution semantics that span machine boundaries. Together, these extensions enable developers to write programs that automatically adapt to different execution scales without explicit modification.

The remainder of this paper is organized as follows: Section~\ref{sec:background} discusses the ubiquity of certain computational concerns and their treatment in language design. Section~\ref{sec:extensions} formalizes our extensions to DSP semantics. Section~\ref{sec:implications} explores the implications of these extensions for different application patterns, and Section~\ref{sec:implementation} presents reflections on the major research challenges and opportunities for the future of scale-agnostic programming.

%% file: 2-casefor.tex
\section{The Case for Scale-Agnostic Language Constructs}
\label{sec:background}

\subsection{Evolution of Intrinsic Language Constructs}
\label{subsec:firstclass}
Over time, we have observed an evolution in what we term \textit{intrinsic language constructs}—fundamental features that are built into the core semantics of a language and handled automatically by its runtime system, rather than requiring explicit manipulation by developers. These constructs represent assumptions and capabilities that are inherent to the language's design philosophy and execution model. Interestingly, some ubiquitous computational concerns have been elevated to intrinsic language construct status despite being fundamentally connected to operating system or hardware capabilities. Python demonstrates this evolution clearly:

\begin{itemize}
    \item \textbf{Memory management:} Python's garbage collection system makes memory management an intrinsic language construct, handling allocation and deallocation automatically. For example, when creating objects like \texttt{message = "Hello, world!"}, Python handles all memory concerns without developer intervention.
    
    \item \textbf{Console I/O:} Despite being a hardware and operating system concern, Python includes built-in functions for console input/output with \texttt{print()} and \texttt{input()}, making these operations first-class citizens in the language specification.
    
    \item \textbf{File operations:} Python provides direct language support for file operations through built-in functions like \texttt{open()} and the context manager pattern (\texttt{with open("filename") as f:}), acknowledging their ubiquity despite being fundamentally operating system services.
\end{itemize}

These examples illustrate that language designers implicitly recognize that certain computational concerns become so ubiquitous that relegating them entirely to libraries creates unnecessary friction. When a concern appears consistently across diverse applications, integrating it directly into language semantics can reduce cognitive overhead and promote more robust implementation patterns.

\subsection{Ubiquitous Yet Overlooked Concerns}
\label{subsec:overlooked}

Several computational concerns have become equally ubiquitous in modern software development but have not yet received comparable treatment as intrinsic language constructs:

\subsubsection{Persistence Across Executions}
\label{subsubsec:persistence}

Nearly all non-trivial applications require some form of state persistence between executions. This need is so fundamental that developers routinely implement persistence mechanisms, whether through databases, file storage, or cloud services. Yet persistence remains largely external to programming language semantics, creating a semantic gap between ephemeral program state and persistent storage.

This separation forces developers to explicitly translate between in-memory representations and persistent formats, increasing complexity and introducing potential inconsistencies. The ubiquity of persistence suggests it should be directly represented in programming models, allowing developers to focus on domain logic rather than persistence mechanics.

\subsubsection{Program Execution Duality}
\label{subsubsec:duality}

The concept of a "program" has evolved significantly since the early days of computing. While traditional batch-processing models assumed a single entry point, single exit point, and linear execution flow, modern applications frequently employ service-oriented architectures with long-running processes and multiple entry points.

This duality of program execution models can be categorized as:

\begin{itemize}
    \item \textbf{Single-State, Single-Entry (SSSE):} The traditional model where a program starts at a designated entry point (e.g., \texttt{main()}), processes input, produces output, and terminates.
    
    \item \textbf{Single-State, Multiple-Entry (SSME):} A long-running program that maintains state between multiple invocations through different entry points (e.g., API endpoints, event handlers).
\end{itemize}

Despite the prevalence of SSME programs in modern software, most programming languages retain a bias toward SSSE semantics, treating SSME patterns as extensions rather than core concerns. This creates a semantic mismatch between language constructs and contemporary application architectures.

\subsubsection{Multi-User Context}
\label{subsubsec:multiuser}

Modern applications typically serve multiple users, each with isolated state and execution flows. This multi-user context is so pervasive that frameworks and platforms have developed extensive machinery to support it, from session management in web frameworks to tenant isolation in cloud services.

However, multi-user semantics rarely appear directly in programming languages. Instead, developers must explicitly model user contexts, manage authentication and authorization, and ensure proper state isolation. This creates unnecessary complexity and increases the risk of security vulnerabilities.

Given the ubiquity of multi-user contexts in contemporary software, programming languages should provide direct mechanisms for modeling user-specific state and behavior. By embedding these concepts in language semantics, developers could write naturally user-aware programs without additional machinery.

\subsubsection{Cross-Machine Distribution}
\label{subsubsec:distribution}

As applications scale, they often need to distribute computation and data across multiple machines. Traditional approaches require developers to explicitly handle distribution concerns such as:

\begin{enumerate}
    \item Data partitioning and sharding
    \item Remote procedure calls
    \item Distributed consistency and consensus
    \item Fault tolerance and recovery
    \item Network topology and latency management
\end{enumerate}

These distribution concerns remain largely external to programming language semantics, leading to significant code changes when scaling from single-machine to distributed deployments. This creates a discontinuity in the development process, often requiring substantial rewrites or architectural shifts when applications outgrow single-machine capacity.

By elevating cross-machine distribution to an intrinsic language construct, programming languages could provide a smoother scaling path, allowing developers to write code that naturally adapts to distributed execution environments without explicit modification.

\subsection{Data Spatial Programming as a Foundation}
\label{subsec:dspfoundation}

Data Spatial Programming~\cite{mars2025dataspatialprogramming} provides an ideal foundation for addressing these overlooked concerns due to its explicit modeling of topological relationships and its inversion of the traditional relationship between data and computation.

In DSP, computation moves to data through walker entities that traverse a graph structure of interconnected nodes. This paradigm naturally models many real-world scenarios, from social networks to agent-based systems.

DSP's archetypes already provide a rich semantic foundation:

\begin{itemize}
    \item \textbf{Nodes} represent discrete locations or entities within a topological structure.
    
    \item \textbf{Edges} represent directed relationships between nodes, encoding both topology and semantics.
    
    \item \textbf{Walkers} represent autonomous computational entities that traverse the node-edge structure.
    
    \item \textbf{Abilities} provide context-sensitive execution triggered by spatial events rather than explicit invocation.
\end{itemize}

These constructs can be extended to represent persistence, multi-user contexts, multiple entry points, and cross-machine distribution as intrinsic aspects of the programming model. By leveraging DSP's topological semantics, we can create a naturally scale-agnostic programming paradigm that spans from individual to enterprise-scale applications without requiring explicit adaptation.

%% file: 3-semantics.tex
\section{Extended DSP Semantics for Scale Agnosticism}
\label{sec:extensions}

We propose three fundamental extensions to DSP semantics that enable scale-agnostic programming: the \textbf{root node} construct, the treatment of \textbf{walkers as entry points}, and \textbf{topology-aware distribution} semantics. Together, these extensions create a programming model where applications naturally adapt to different scales without explicit modification.

\subsection{The Root Node: A Persistent Anchor in Topological Space}
\label{subsec:rootnode}

We introduce the concept of a \textbf{root node}, a special node instance that serves as a persistent anchor within the program's topological space. The root node differs from standard nodes in several important ways:

\begin{enumerate}
    \item \textbf{Global Accessibility:} The root node is accessible from anywhere in the program through a special keyword \texttt{root}, similar to existing keywords like \texttt{this} and \texttt{self} common to languages like C++/Java and Python respectively, and \texttt{here} in DSP, but with global scope.
    
    \item \textbf{Automatic Persistence:} The root node and any nodes connected to it (directly or indirectly) automatically persist across program executions. The runtime system manages this persistence without explicit developer intervention.
    
    \item \textbf{User Association:} Each user interacting with the program has a distinct root node, creating natural isolation between user contexts. The runtime system automatically associates incoming requests with the appropriate user's root node.
\end{enumerate}

Formally, we define the root node as a special instance of a node class:

\[
\texttt{root} \in \tau_{\text{node}}
\]

with the additional property that for any program $P$ and user $u$:

\[
\texttt{root}_u^P \text{ persists across all executions of } P \text{ for user } u
\]

Furthermore, for any node $n$ reachable from the root node through a path of edges:

\[
\exists \text{ path } (e_1, e_2, \ldots, e_k) \text{ where } e_1 = (\texttt{root}, n_1, d_1), \ldots, e_k = (n_{k-1}, n, d_k) \Rightarrow n \text{ persists}
\]

This creates a \textit{persistence by reachability} model where developers can simply connect nodes to the root (directly or indirectly) to make them persistent, without special persistence annotations or explicit storage operations.

\subsubsection{Topological Persistence}
\label{subsubsec:topologicalpersistence}

The root node acts as an anchor for a subgraph of the program's overall topology. Any node reachable from the root node (through edges) becomes part of this persistent subgraph and is automatically preserved between program executions.

This creates a natural boundary between ephemeral and persistent state within the program. Elements connected to the root persist, while disconnected elements remain transient. Developers can control persistence simply by managing connectivity to the root node, without separate persistence mechanisms.

Formally, let $G_{\text{root}} = (N_{\text{root}}, E_{\text{root}})$ be the subgraph reachable from the root node, where:

\begin{align}
N_{\text{root}} &= \{\texttt{root}\} \cup \{n \in \tau_{\text{node}} \mid \exists \text{ path from \texttt{root} to } n\} \\
E_{\text{root}} &= \{e \in \tau_{\text{edge}} \mid e = (n_{\text{src}}, n_{\text{dst}}, d) \text{ where } n_{\text{src}}, n_{\text{dst}} \in N_{\text{root}}\}
\end{align}

Then the persistence property ensures:

\[
G_{\text{root}}^{t} \text{ is preserved in } G_{\text{root}}^{t+1}
\]

where $G_{\text{root}}^{t}$ represents the state of the persistent subgraph at execution time $t$.

\subsubsection{Multi-User Context Through Root Isolation}
\label{subsubsec:rootmultiuser}

Each user interacting with the program has a distinct root node, creating automatic isolation between user contexts. This leads to a collection of user-specific persistent subgraphs, each anchored to a different root node instance.

Formally, for users $u_1$ and $u_2$:

\[
\texttt{root}_{u_1} \neq \texttt{root}_{u_2}
\]

And consequently:

\[
G_{\text{root}_{u_1}} \cap G_{\text{root}_{u_2}} = \emptyset
\]

This isolation happens automatically at the language level, without requiring explicit session management or user context handling. Developers simply write programs relative to the root node, and the runtime system ensures proper isolation between users.

\subsection{Walkers as Entry Points: From Traversals to APIs}
\label{subsec:walkerentry}

Our second key extension treats walkers as potential entry points into the program, enabling Single-State, Multiple-Entry (SSME) execution models. Rather than requiring a single designated entry point, any walker can serve as an entry point, with the runtime system managing activation based on external invocations.

\subsubsection{Entry Point Registration}
\label{subsubsec:entrypoints}

Walkers can be designated as entry points through an entry point annotation or declaration. When a walker is designated as an entry point, the runtime system automatically generates appropriate interfaces for external invocation, such as API endpoints, event handlers, or command-line interfaces.

For a walker class $W \in \tau_{\text{walker}}$, the entry point designation is expressed as:

\[
\text{EntryPoint}(W) \Rightarrow W \text{ becomes an external invocation target}
\]

When the entry point is invoked externally, the runtime system:

\begin{enumerate}
    \item Instantiates a new instance of the walker: $w = \text{new } W()$
    
    \item Maps invocation parameters to walker properties: $w.p_i = \text{param}_i$
    
    \item Spawns the walker at the user's root node: $w \Rightarrow \texttt{root}$
    
    \item Returns results after traversal completes: $\text{result} = w.\text{result}$
\end{enumerate}

This process transforms walker traversals into API-like interfaces, allowing external systems to invoke walker behaviors within the program's topological space.

\subsubsection{Parameter and Result Mapping}
\label{subsubsec:paramapping}

To support external invocation, walker entry points include parameter mapping specifications that determine how external parameters map to walker properties and how walker results are returned to callers.

For a walker entry point $W$, the parameter mapping includes:

\begin{itemize}
    \item \textbf{Input Parameter Mapping:} Specifications for how invocation parameters map to walker properties, potentially including type conversions and validation.
    
    \item \textbf{Result Mapping:} Specifications for which walker properties should be returned as results, including transformation and formatting options.
\end{itemize}

This mapping enables walkers to serve as clean interfaces between external invocations and internal traversal behaviors, without requiring additional adapter layers.

\subsubsection{Execution Context Management}
\label{subsubsec:execcontext}

When a walker entry point is invoked, the runtime system automatically establishes the appropriate execution context based on the invoking user or system. This includes:

\begin{enumerate}
    \item Identifying the appropriate user context based on authentication information
    
    \item Locating or creating the user's root node
    
    \item Setting up isolation boundaries to prevent cross-user data access
    
    \item Establishing transaction or consistency boundaries as needed
\end{enumerate}

This automatic context management ensures that invocations from different users remain properly isolated, even in concurrent execution environments.

\subsection{Topology-Aware Distribution: Spanning Machine Boundaries}
\label{subsec:distribution}

Our third key extension adds distribution semantics to DSP, allowing programs to naturally span multiple machines without explicit adaptation. This extension leverages the topological nature of DSP to create a seamless distribution model. We present two fundamental strategies for distribution in DSP environments: data-centric distribution and computation-centric distribution.

\subsubsection{Data-Centric Distribution}
\label{subsubsec:datacentric}

In data-centric distribution, the focus is on optimally placing nodes across the distributed infrastructure while keeping walkers relatively stationary. This approach mirrors traditional distributed database systems like Redis or distributed key-value stores, where data partitioning is the primary distribution mechanism.

The placement of nodes across machines follows a set of principles that balance performance, data locality, and resource utilization:

\begin{equation}
\text{Placement}(n) = f(\text{AccessPat}(n), \text{DataVol}(n), \text{ResAvail})
\end{equation}

Where $f$ represents the placement function that maps nodes to physical machines based on observed access patterns, data volume, and available resources. This function can be implemented through various strategies, from static partitioning schemes to dynamic, adaptive approaches that evolve based on runtime behavior.

For nodes $n_i$ and $n_j$ with strong topological or semantic relationships, the placement function aims to maximize co-location probability:

\begin{align}
P(&\text{SameMachine}(n_i, n_j) | \text{Related}(n_i, n_j)) > \nonumber \\
&P(\text{SameMachine}(n_i, n_k) | \neg\text{Related}(n_i, n_k))
\end{align}

This preferential co-location of related nodes minimizes cross-machine traversals, reducing latency and network overhead. The runtime system may implement this through techniques such as graph partitioning algorithms that analyze the topology to identify clusters of highly interconnected nodes.

In data-centric distribution, when a walker needs to access a node on a remote machine, the system employs remote data access protocols:

\begin{equation}
\text{Access}(w, n) = 
\begin{cases}
\text{LocalAccess}(w, n) & \text{if } M(w) = M(n) \\
\text{RemoteAccess}(w, n) & \text{otherwise}
\end{cases}
\end{equation}

Remote access operations may involve synchronous requests for strong consistency guarantees or asynchronous operations for eventual consistency models. The choice of consistency model affects both the performance characteristics and the programming semantics:

\begin{equation}
\text{Latency}(\text{Access}(w, n)) \propto \text{ConsistencyStrength}(M)
\end{equation}

Data-centric distribution is particularly well-suited for applications with high node-to-walker ratios, where nodes represent substantial data volumes, or where walker computation is lightweight compared to the cost of data transfer.

\subsubsection{Computation-Centric Distribution}
\label{subsubsec:computationcentric}

In computation-centric distribution, walkers actively migrate between machines to co-locate with the nodes they need to access. This approach inverts the traditional distributed computing paradigm by moving computation to data rather than data to computation, aligning perfectly with DSP's core philosophy.

When a walker traverses an edge that crosses machine boundaries, the runtime system initiates a migration process:

\begin{align}
\text{Traverse}(w, e = (n_s, n_d, l)) = 
\begin{cases}
\text{LocalTrav}(w, e) & \text{if } M(n_s) = M(n_d) \\
\text{Migrate}(w, M(n_d)) + \text{LocalTrav}(w, e) & \text{otherwise}
\end{cases}
\end{align}

The migration process involves serializing the walker's state, transferring it to the target machine, and reconstituting the walker to continue execution:

\begin{equation}
\text{Migrate}(w, m) = \text{Deser}_m(\text{Transfer}(\text{Ser}(w)))
\end{equation}

This approach maintains execution locality by ensuring that walkers always execute on the same machine as the nodes they're currently traversing. This locality preserves the semantics of node-walker interactions and eliminates the need for remote procedure calls or distributed synchronization mechanisms within the application logic.

The migration process is entirely transparent to the walker implementation. From the walker's perspective, traversal semantics remain consistent whether crossing local or remote edges:

\begin{equation}
\forall e \in \tau_{\text{edge}} : \text{Semantics}(\text{Traverse}(w, e)) \text{ invariant to } M(e.s) = M(e.d)
\end{equation}

This transparency extends to the handling of walker state during migration. The runtime system automatically manages the serialization of both the walker's explicit properties and its execution context, including local variables, continuation points, and runtime state:

\begin{equation}
\text{Ser}(w) = \text{SerProps}(w) \cup \text{SerExecCtx}(w)
\end{equation}

Computation-centric distribution proves particularly effective for applications with complex walker logic, relatively small walker state footprints, or where data locality significantly impacts performance. It also naturally accommodates heterogeneous computing environments, as walkers can utilize specialized hardware capabilities available on different machines as they migrate through the system.

\subsubsection{Hybrid Distribution Models}
\label{subsubsec:hybridmodels}

In practice, optimal distribution often involves a hybrid approach that combines elements of both data-centric and computation-centric strategies. The runtime system can make dynamic decisions about whether to move data to computation or computation to data based on factors such as:

\begin{align}
\text{Strategy}(w, n) = 
\begin{cases}
\text{MoveData} & \text{if } S(n) \ll S(w) \text{ or} \\
 & \text{AccessPat}(n) \text{ highly localized} \\
\text{MoveComp} & \text{if } S(w) \ll S(n) \text{ or} \\
 & \text{AccessPat}(w) \text{ forms clear path}
\end{cases}
\end{align}

These hybrid models can be influenced by developer-provided affinity hints that express knowledge about expected access patterns or relative sizes that might not be apparent to the runtime system through static analysis alone.

Regardless of the specific distribution strategy employed, the key benefit of our approach is that distribution remains transparent to application code. Developers write programs using the same topological semantics regardless of whether the application runs on a single machine or spans a distributed cluster. The runtime system assumes responsibility for implementing the chosen distribution strategy, managing consistency, and handling the complexities of cross-machine communication.

\subsubsection{Fault Tolerance and Recovery}
\label{subsubsec:faulttolerance}

Distribution inherently introduces the possibility of partial system failures, where some machines become unavailable while others continue functioning. Our extended DSP semantics include built-in mechanisms for handling such failures transparently to application code.

For data-centric distribution, node replication provides the primary fault tolerance mechanism:

\begin{equation}
\forall n \in \tau_{\text{node}} : \text{Replicas}(n) = \{n_1, n_2, \ldots, n_k\} \text{ where } k \geq \text{RepFactor}(n)
\end{equation}

The runtime system manages consensus among these replicas, potentially using algorithms like Paxos or Raft for strong consistency guarantees, or simpler eventual consistency mechanisms for less critical data.

For computation-centric distribution, walker checkpointing provides resilience against machine failures:

\begin{equation}
\text{Checkpoint}(w, t) = \text{Persist}(\text{Ser}(w, t))
\end{equation}

Where $t$ represents a specific point in the walker's execution. After a failure, walkers can be reconstituted from their most recent checkpoint:

\begin{equation}
\text{Recover}(w) = \text{Deser}(\text{LatestCheckpoint}(w))
\end{equation}

These fault tolerance mechanisms operate transparently to application code, allowing developers to write programs as if machine failures never occurred, while the runtime system handles recovery behind the scenes. This approach maintains the scale-agnostic nature of DSP even in the face of distributed system failures, preserving the programming model's simplicity while providing the robustness required for production deployments.

\subsection{Combining Root, Walker, and Distribution Extensions}
\label{subsec:combining}

When combined, the root node, walker entry point, and distribution extensions create a powerful foundation for scale-agnostic programming. Developers can write programs that:

\begin{enumerate}
    \item Naturally persist state by connecting nodes to the root
    
    \item Automatically isolate user contexts through user-specific root nodes
    
    \item Expose functionality through walker entry points without additional API layers
    
    \item Span multiple machines transparently without explicit distribution code
    
    \item Adapt to different scales without explicit modification
\end{enumerate}

This integration bridges the gap between topological semantics and operational concerns, allowing developers to focus on domain logic while delegating runtime mechanics to the execution environment.

%% file: 4-implications.tex
\section{Implications for Application Patterns}
\label{sec:implications}

The extensions to DSP semantics presented in this paper have profound implications for how developers design and implement applications. By providing intrinsic language constructs for persistence, multi-user contexts, multiple entry points, and distribution, these extensions enable new patterns that naturally scale across different execution contexts. In this section, we examine several key application patterns that benefit from our scale-agnostic approach, discussing how each represents a significant advancement over traditional programming models.

\subsection{Natural Transition from Single-User to Multi-User}
\label{subsec:singletomulit}

Perhaps the most significant impact of our extensions is the seamless transition from single-user to multi-user applications. In traditional programming paradigms, adapting a single-user application for multiple users necessitates extensive architectural modifications. Developers must implement user authentication and authorization frameworks, establish session management systems, partition data according to user identities, ensure isolation between user contexts, and introduce concurrency controls for shared resources. This transformation typically requires substantial codebase alterations, often leading to complete architectural redesigns.

With our extended DSP semantics, these concerns are addressed intrinsically within the language and runtime system. The automatic allocation of distinct root nodes to each user creates natural partitioning of state based on connectivity to user-specific roots. The runtime system maintains isolation between these partitions while automatically managing concurrency for walker executions across user contexts. This approach provides a form of automatic multi-tenancy that preserves application semantics across usage scales.

Consider the implementation of a document editing application. In traditional paradigms, transitioning from a single-user local editor to a collaborative multi-user system requires fundamental architectural changes. With our approach, a developer simply models documents as nodes connected to the user's root node, and collaborative operations as walker traversals. The same code operates correctly whether deployed as a personal application or scaled to enterprise usage, with the runtime system handling the complexity of user isolation and concurrent operations.

\subsection{Service-Oriented Architectures Without Implementation Complexity}
\label{subsec:soaboilerplate}

Service-oriented architectures have become ubiquitous in modern application development, yet they typically demand extensive implementation overhead. Developers must define service interfaces, implement request handling logic, manage serialization and deserialization of parameters, establish routing mechanisms, and implement comprehensive error-handling protocols. This infrastructure code often overshadows the actual service logic, increasing development complexity and maintenance burden.

Our walker-as-entry-point extension eliminates much of this complexity by making service interfaces an intrinsic part of the programming model. Walker classes directly represent service behaviors, with their definition implicitly establishing the service interface. Parameter mappings handle serialization and deserialization automatically, while the spawn operation manages routing to the correct user context. Traversal semantics naturally express complex process flows without requiring explicit orchestration mechanisms.

This reduction in implementation complexity allows developers to focus on domain logic rather than service mechanics. A complex microservice architecture can be expressed directly in terms of the relevant topological constructs—walkers, nodes, and edges—without additional adaptation layers or infrastructure code. The result is a more direct mapping between the conceptual service architecture and its implementation, reducing both cognitive load and development effort.

\subsection{Persistence Without Representational Impedance}
\label{subsec:persistence}

Persistence in traditional application development creates significant friction due to the impedance mismatch between in-memory and persistent data representations. Developers typically need to define database schemas, implement object-relational mappings, manage serialization and deserialization processes, and explicitly orchestrate state synchronization between memory and storage. These mechanisms add considerable complexity to application code and introduce potential inconsistencies between different state representations.

Our root node extension transforms persistence into a natural consequence of topological connectivity within the application graph. Any node connected to the root node—directly or indirectly—automatically persists across program executions. This approach eliminates the need for explicit serialization or separate database schema definitions. The in-memory object graph and its persistent representation become conceptually unified, removing the impedance mismatch that has traditionally complicated state management.

This topological persistence model enables a more natural expression of domain models, where relationships between entities are represented directly as edges between nodes. Developers can work with connected object graphs without worrying about how they map to underlying storage mechanisms, creating a more coherent relationship between domain concepts and their implementation. The persistence boundary becomes a natural topological property—connected to root means persistent, disconnected means transient—rather than requiring explicit persistence annotations or frameworks.

\subsection{Distributed Execution Without Distribution Semantics}
\label{subsec:distributionsystems}

Distributed systems development traditionally requires developers to explicitly address numerous concerns related to cross-machine execution. These include data partitioning strategies, remote procedure call mechanisms, consistency protocols, and fault tolerance systems. This additional code significantly increases implementation complexity and often leads to architectural designs that are tightly coupled to specific distribution patterns.

With our distribution extension, these concerns are handled automatically by the runtime system without requiring explicit distribution semantics in application code. Node placement across machines is managed transparently, while walker migration occurs automatically during edge traversals. Consistency models can be configured without modifying application logic, and fault tolerance mechanisms operate behind the scenes. This separation between application semantics and distribution mechanics enables true scale-agnostic programming.

The implications of this approach are particularly profound for applications that need to scale dynamically based on user load or data volume. A program written for single-machine execution can automatically scale out to a distributed cluster without modification as requirements evolve. The distribution semantics ensure consistent behavior across machine boundaries, preserving the program's logical semantics regardless of its physical execution environment. This characteristic substantially reduces the development and operational complexity associated with building scalable systems.

\subsection{Cohesive State Management Across Execution Contexts}
\label{subsec:statefulapps}

State management in traditional application development introduces significant complexity, particularly in distributed environments. Developers must implement mechanisms for state initialization, tracking changes, ensuring consistency, persisting modifications, and synchronizing state across distributed components. These concerns often lead to complex state management patterns that are difficult to reason about and maintain.

Our extended DSP semantics provides a more cohesive approach to state management through its unification of in-memory state, persistence, and distribution. State is naturally represented as a graph of nodes connected to the user's root node. Updates flow through this graph via walker traversals, with changes persisting automatically without explicit save operations. The runtime system maintains consistency according to the configured model, and state can span machine boundaries without requiring distribution-specific code.

This approach simplifies the development of stateful applications by reducing the cognitive overhead associated with state management. The topological representation of state aligns more naturally with domain models, while the automatic handling of persistence and distribution concerns allows developers to focus on business logic. The result is a more direct mapping between conceptual state transitions and their implementation, creating applications that are more robust and easier to maintain across different execution scales.

\subsection{Unified Programming Model Across Application Scales}
\label{subsec:unifiedmodel}

Perhaps the most significant implication of our extended DSP semantics is the establishment of a unified programming model that spans different application scales. Traditional development approaches often require different programming models, frameworks, and patterns depending on whether an application is single-user or multi-user, stateless or stateful, local or distributed. This fragmentation increases the cognitive load on developers and creates artificial boundaries between application categories.

Our scale-agnostic approach eliminates these boundaries by providing a consistent set of abstractions that work seamlessly across scales. The same programming model applies whether developing a personal utility, a departmental application, or an enterprise-scale system. This continuity allows developers to leverage their knowledge and code consistently across projects of different scales, reducing the learning curve associated with scale transitions.

Furthermore, this unified model enables more natural application evolution. A successful personal application can grow into a multi-user system and eventually into a distributed enterprise application without fundamental redesigns. The same core code can scale from prototype to production, with the runtime system assuming responsibility for the operational concerns associated with larger scales. This characteristic not only improves developer productivity but also enables more sustainable application lifecycles where successful prototypes can evolve directly into production systems rather than requiring complete rewrites.

%% file: 5-implementation.tex
\section{Research Challenges and Opportunities}
\label{sec:implementation}

While our semantic extensions provide powerful abstractions for scale-agnostic programming, our current implementation in Jaseci~\cite{jasecical,JaseciGitHub} uses only naive heuristics for runtime functionality. A robust implementation requires careful consideration of several key research challenges. In this section, we discuss research objectives across different dimensions of the scale-agnostic model, emphasizing the importance of environmental adaptivity in the runtime system. A central theme in our approach is that the runtime should automatically detect execution context characteristics and configure implementation mechanisms accordingly, allowing applications to seamlessly transition between different operational environments without explicit reconfiguration.

\subsection{Persistence Strategies}
\label{subsec:impl_persistence}

The root node's persistence guarantee represents a significant implementation challenge that necessitates flexible storage strategies adaptable to different deployment contexts. A key insight in our approach is that persistence mechanisms should be automatically selected and configured by the runtime system based on contextual factors including deployment scale, performance requirements, data characteristics, and available infrastructure.

\paragraph{\textbf{Database Mapping}} -- The most conventional implementation approach automatically maps the persistent subgraph to an underlying database schema. Unlike traditional object-relational mapping frameworks that require explicit mapping definitions, our approach derives mapping patterns directly from topological connectivity. Nodes become entities, edges become relationships, and properties become attributes. The runtime system can analyze the graph structure to automatically generate and evolve database schemas, potentially selecting different database technologies based on the topological characteristics of the persistent subgraph. For instance, highly connected graphs might leverage graph databases, while more hierarchical structures might utilize document stores.

\paragraph{\textbf{Serialization-Based Persistence}} -- This alternative approach is particularly suited to smaller deployments or development environments. In this model, the persistent subgraph is serialized to a structured format such as JSON, XML, or a custom binary representation. For efficiency, the runtime can implement incremental serialization, persisting only modified portions of the graph between execution cycles. This approach can be augmented with journaling mechanisms that record operations rather than states, enabling more efficient storage and robust recovery capabilities. The runtime system should automatically select serialization formats and strategies based on the graph's size, update patterns, and performance characteristics of the deployment environment.

\paragraph{\textbf{Specialized Object Stores}} -- Graph databases, multi-model databases, or distributed key-value stores with relationship support offer a third implementation strategy that aligns naturally with DSP's topological semantics. These systems can directly represent the persistent subgraph with minimal transformation and often provide specialized query capabilities and traversal optimizations that complement DSP's walker-based computation model. The runtime system can automatically map node and edge patterns to the native constructs of these specialized stores, selecting appropriate technologies based on the application's scale and access patterns.

\paragraph{\textbf{In-Memory with Durability}} -- For deployment contexts with stringent performance requirements, this approach offers the highest performance profile. It maintains the persistent subgraph primarily in memory, with background processes ensuring durability through techniques such as write-ahead logging, journaling, or periodic snapshots. Similar to systems like Redis or Apache Ignite, this approach prioritizes access speed while still providing durability guarantees. The runtime should automatically adopt this strategy when deployment contexts indicate high-performance requirements, potentially scaling from single-instance to distributed in-memory clusters as application load increases.

\paragraph{\textbf{Adaptive Strategy Selection}} -- The key notion is that the runtime system should detect environmental characteristics and seamlessly transition between these persistence strategies as applications scale. A personal application might begin with simple serialization, transition to database mapping as user base grows, and eventually leverage distributed object stores or in-memory persistence with durability for enterprise-scale deployments—all without requiring application code changes. This capability embodies the essence of our scale-agnostic approach, where implementation details adapt to scale while preserving consistent semantics across execution contexts.

\subsection{User Context Resolution}
\label{subsec:impl_usercontext}

Associating incoming requests with the correct user-specific root node requires sophisticated context resolution mechanisms that integrate with authentication and identity management systems. The runtime system must automatically detect available identity infrastructure in the deployment environment and configure appropriate resolution strategies.

\paragraph{\textbf{Authentication Integration}} -- This forms the foundation of user context resolution, requiring the runtime system to interface with diverse authentication providers. In simple deployments, this might involve basic username/password validation against local stores, while complex enterprise environments might require integration with LDAP directories, OAuth providers, OpenID Connect systems, or SAML identity providers. The runtime should detect available authentication infrastructure and automatically configure integration points, mapping authenticated identities to specific root nodes. This integration should leverage existing authentication mechanisms rather than creating parallel systems, allowing applications to benefit from established security practices in their deployment environments.

\paragraph{\textbf{Session Management}} -- This extends authentication by maintaining user context across multiple interactions, particularly in interactive applications. The runtime system must integrate session tracking with root node resolution to ensure consistent access to the appropriate persistent subgraph. This integration varies significantly across deployment contexts—web applications might leverage cookies or tokens, mobile applications might maintain device-specific sessions, and enterprise systems might employ federated session management. Rather than requiring explicit session handling in application code, the runtime should automatically detect the execution environment and implement appropriate session mechanisms, ensuring contextual continuity while adapting to each environment's security constraints and operational patterns.

\paragraph{\textbf{Cross-cutting Identity}} -- This represents a particularly complex challenge in enterprise environments where users may have identities spanning multiple systems, potentially with different representations across business units or functional domains. In these contexts, the runtime must implement federated identity resolution, potentially mapping multiple identity tokens to a single conceptual user and their associated root node. This resolution might involve identity attribute aggregation, role normalization, or hierarchical identity mapping. The runtime should automatically detect these complex identity landscapes and configure appropriate resolution strategies, shielding applications from the underlying complexity while ensuring consistent user context association.

\paragraph{\textbf{Adaptive Identity Integration}} -- The key insight in our approach to user context resolution is that the runtime system should automatically adapt to the identity infrastructure of each deployment environment. An application might use simple local authentication during development, standard web authentication in initial deployment, and enterprise identity integration at scale—all without requiring code changes. This adaptivity enables applications to seamlessly transition across deployment contexts while maintaining consistent semantics for user isolation and context management.

\subsection{Entry Point Exposure}
\label{subsec:impl_entrypoints}

Walker entry points represent the interfaces through which external systems interact with DSP applications. The runtime system must expose these entry points through appropriate mechanisms based on the deployment context, automatically detecting environmental characteristics and configuring interface adaptations accordingly.

\paragraph{\textbf{REST API Exposure}} -- In web deployment contexts, the runtime system should automatically expose walker entry points as REST APIs, mapping walker parameters to request formats and walker results to response structures. This mapping should follow established REST conventions, including appropriate HTTP method selection based on walker semantics (GET for read operations, POST for modifications, etc.), status code mapping based on execution outcomes, and content negotiation for flexible result formats. The runtime might generate OpenAPI specifications automatically from walker definitions, enabling integration with API management platforms and developer tooling. For web contexts with different architectural patterns, the runtime might alternatively expose walkers through GraphQL interfaces or WebSocket endpoints, depending on detected environmental patterns and application characteristics.

\paragraph{\textbf{RPC Interface Generation}} -- For distributed system deployments, particularly in enterprise environments, RPC interfaces often provide more appropriate interaction patterns. The runtime should detect distributed deployment contexts and automatically expose walker entry points as RPC methods using protocols like gRPC, Apache Thrift, or other available communication frameworks. This exposure includes generating appropriate interface definitions (such as Protocol Buffers or IDL files), implementing serialization for complex parameter and result types, and managing connection pooling and reliability patterns. The runtime should select specific RPC technologies based on detected infrastructure and deployment patterns, allowing applications to leverage existing communication frameworks without explicit adaptation.

\paragraph{\textbf{Event Handler Registration}} -- Event-driven architectures represent another common deployment pattern, particularly for asynchronous or reactive systems. In these contexts, the runtime should automatically register walker entry points as event handlers with appropriate routing and filtering rules. This registration might involve integration with message brokers like Apache Kafka, RabbitMQ, or cloud-native event systems, with the runtime automatically mapping event payloads to walker parameters and potentially publishing result events. The runtime should detect the available event infrastructure and configure appropriate integration patterns, potentially implementing event sourcing, CQRS patterns, or other event-driven architectural styles depending on the deployment context.

\paragraph{\textbf{Command-Line Interface Generation}} -- For local execution environments, including development scenarios, command-line interfaces often provide the most appropriate interaction model. The runtime should detect local execution contexts and automatically expose walker entry points as command-line operations with generated argument parsing, help documentation, and output formatting. This exposure should follow platform-specific conventions, potentially integrating with shell completion systems and local tooling ecosystems. As applications scale to server deployment, the runtime should seamlessly transition from these local interfaces to network-accessible APIs without requiring code modifications.

\paragraph{\textbf{Context-Aware Interface Adaptation}} -- The unifying principle across these exposure mechanisms is that the runtime system should automatically detect the deployment context and configure appropriate interface adaptations. An application might expose command-line interfaces during development, REST APIs in web deployment, and a combination of RPC interfaces and event handlers in enterprise settings—all without requiring changes to the underlying walker implementations. This capability enables applications to maintain consistent semantics while adapting their interaction patterns to each operational environment, embodying the essence of scale-agnostic programming.

\subsection{Distribution Optimization}
\label{subsec:impl_distribution}

Implementing efficient distribution semantics across machine boundaries presents perhaps the most challenging aspect of our scale-agnostic model. The runtime system must automatically detect distribution requirements and configure appropriate mechanisms to maintain programming model semantics while optimizing for performance, reliability, and resource utilization.

\paragraph{\textbf{Topology-aware Partitioning}} -- This serves as the foundation for effective distribution, requiring the runtime to analyze the application's node-edge structure and identify natural partition boundaries. This analysis should consider both static topology and dynamic access patterns, potentially employing graph partitioning algorithms that minimize cross-partition edges while balancing load across machines. As applications scale, the runtime should continuously monitor traversal patterns and adaptively refine partitioning strategies, potentially migrating nodes between machines to optimize locality based on observed walker behaviors. This adaptive partitioning should happen transparently to application code, maintaining consistent semantics while optimizing underlying distribution patterns.

\paragraph{\textbf{Communication Optimization}} -- Network efficiency across partition boundaries significantly impacts distributed performance, requiring sophisticated strategies for reducing overhead. The runtime should automatically select appropriate communication patterns based on detected infrastructure capabilities and application characteristics. These patterns might include batched updates to reduce message counts, compression for bandwidth reduction, delta encoding for efficient state synchronization, or predictive prefetching based on likely traversal paths. For walker migration across machines, the runtime should optimize serialization formats and transfer protocols to reduce migration latency, potentially implementing specialized formats for common walker patterns. These optimizations should adapt to observed network characteristics and available infrastructure without requiring application awareness.

\paragraph{\textbf{Consistency Management}} -- Balancing performance and semantic guarantees across distributed nodes requires careful management. The runtime should automatically select appropriate consistency mechanisms based on application patterns and deployment requirements, potentially employing different strategies for different regions of the node-edge graph. These strategies might include optimistic concurrency with conflict resolution for high-contention scenarios, causal consistency for naturally related operations, or eventual consistency with background synchronization for loosely-coupled components. The runtime should detect natural consistency boundaries within applications and apply the most efficient consistency model that preserves required semantics, adapting these models as applications scale across machine boundaries.

\paragraph{\textbf{Fault Tolerance Integration}} -- Handling failures in distributed environments introduces significant complexity that must be managed transparently by the runtime system. The runtime should automatically implement appropriate resilience patterns based on detected infrastructure reliability and application criticality. These patterns might include node replication for high-availability, walker checkpointing for recoverability, partition-aware retry mechanisms, or compensating transactions for distributed rollback capabilities. As applications scale to larger clusters, the runtime should automatically increase resilience mechanisms proportionally to the increased failure probability, potentially implementing sophisticated consensus protocols for critical coordination points. These mechanisms should operate without application awareness, maintaining consistent fault behavior semantics across deployment scales.

\paragraph{\textbf{Scale-Adaptive Distribution}} -- The central insight in our approach to distribution is that the runtime system should automatically detect scale requirements and infrastructure characteristics, adapting distribution strategies accordingly. An application might execute entirely on a single machine during development, scale to a small cluster for initial deployment, and eventually span massive distributed infrastructure—all while maintaining consistent programming semantics and without requiring code modifications. This adaptivity enables true scale-agnostic programming, where applications can seamlessly transition across deployment scales while preserving their fundamental behavior and developer mental model.